\documentclass[twocolumn, showpacs, superscriptaddress, nofootinbib]{revtex4-1}
\usepackage[utf8]{inputenc}
\usepackage[english]{babel}
\usepackage{graphicx}
\usepackage{float}
\usepackage{amsmath}
\usepackage{bm}
\usepackage{environ}
\usepackage{comment}
\usepackage{grffile}
\usepackage{xcolor}

\usepackage{booktabs}       % for nice tables
\usepackage{placeins}       % floatbarrier
\usepackage{amsfonts}
\usepackage{amssymb}
\usepackage{wasysym}
\usepackage{dsfont}
\usepackage{color,soul}
\usepackage{dcolumn}		% Align table columns on decimal point
\usepackage{bm}			    % bold math
\usepackage[pdftex, colorlinks, citecolor=blue, linkcolor=blue, urlcolor=blue]{hyperref}
\usepackage[all]{hypcap}	% makes hyperref work OK
\usepackage{enumitem}   
\graphicspath{{./figures/}}

\begin{document}
\title{Effective description of bistability and irreversibility in apoptosis}
\author{Sol M. Fern\'andez Arancibia} 
\affiliation{Instituto de Investigaci\'on en Biomedicina de Buenos Aires (IBioBA) – CONICET/Partner Institute of the Max Planck Society, Polo Cient\'{i}fico Tecnol\'ogico, Godoy Cruz 2390, Buenos Aires C1425FQD, Argentina}
\author{Hern\'an E. Grecco}		
% \affiliation{Departamento de F\'{\i}sica, FCEyN UBA, Ciudad Universitaria, 1428 Buenos Aires, Argentina}%
% \affiliation{IFIBA, CONICET, Buenos Aires, Argentina}
% \affiliation{Department of Physics, FCEN, University of Buenos Aires and IFIBA, CONICET, Buenos Aires, Argentina} %% HERNAN
\affiliation{Department of Physics, FCEN, University of Buenos Aires, Ciudad Universitaria, 1428 Buenos Aires, Argentina} %% HERNAN
\affiliation{IFIBA, CONICET, Buenos Aires, Argentina} %% HERNAN
\affiliation{Max Planck Institute for Molecular Physiology, Department of Systemic Cell Biology, Otto-Hahn-Strasse~11, Dortmund D-44227, Germany}
\author{Luis G. Morelli}		
\affiliation{Instituto de Investigaci\'on en Biomedicina de Buenos Aires (IBioBA) – CONICET/Partner Institute of the Max Planck Society, Polo Cient\'{i}fico Tecnol\'ogico, Godoy Cruz 2390, Buenos Aires C1425FQD, Argentina}
% \affiliation{Departamento de F\'{\i}sica, FCEyN UBA, Ciudad Universitaria, 1428 Buenos Aires, Argentina}%
\affiliation{Department of Physics, FCEN, University of Buenos Aires, Ciudad Universitaria, 1428 Buenos Aires, Argentina} %% HERNAN
\affiliation{Max Planck Institute for Molecular Physiology, Department of Systemic Cell Biology, Otto-Hahn-Strasse~11, Dortmund D-44227, Germany}
%\affiliation{Department of Systemic Cell Biology, Max Planck Institute for Molecular Physiology, Otto-Hahn-Str.~11, 44227, Dortmund, Germany}

\date{\today}
\begin{abstract}
Apoptosis is a mechanism of programmed cell death in which cells engage in a controlled demolition and prepare to be digested without damaging their environment. In normal conditions apoptosis is repressed, until it is irreversibly induced by an appropriate signal. In adult organisms apoptosis is a natural way to dispose of damaged cells, and its disruption or excess is associated with cancer and autoimmune diseases. Apoptosis is regulated by a complex signaling network controlled by caspases, specialized enzymes that digest essential cellular components and promote the degradation of genomic DNA. In this work we propose an effective description of the signaling network focused on caspase-3 as a readout of cell fate. We integrate intermediate network interactions into a nonlinear feedback function acting on caspase-3 and introduce the effect of pro-apoptotic stimuli and regulatory elements as a saturating activation function. We find that the theory has a robust bistable regime where two possible states coexist, representing survival and cell death fates. For a broad range of parameters, strong stimuli can induce an irreversible switch to the death fate. We use the theory to explore dynamical stimulation conditions and determine how cell fate depends on different stimuli patterns. This analysis reveals a critical relation between transient stimuli intensity and duration to trigger irreversible apoptosis.
\end{abstract}
\maketitle

\section{Introduction}
Cells have intrinsic control mechanisms that can trigger cell death programs, such as apoptosis~\cite{bedoui20}.
During apoptosis, cells undergo an orderly demolition to avoid inflammation in the surrounding tissues~\cite{alberts02}. 
Apoptosis plays key roles during embryonic development, shaping tissues~\cite{meier00, ambrosini17}, and in adult organisms, removing damaged cells~\cite{henson01} and controlling tissue homeostasis~\cite{bedoui20}. 
Impairment or malfunction of apoptosis is associated with a wide range of developmental abnormalities and diseases such as cancer and autoimmune disorders~\cite{favaloro12, singh19}.

Apoptosis is regulated by a signaling network composed of specialized enzymes termed caspases, that cleave essential components of the cell and promote DNA degradation~\cite{alberts02, thornberry97, stennicke00}. 
Inactive caspase precursors termed procaspases, lack the capacity to cleave their substrates.
Procaspases are conformed by two subunits and become active caspases by cleavage of the inter subunit linker~\cite{julien17}.
Caspases involved in apoptosis can be broadly classified in two groups: initiators caspase-8 and caspase-9, and effectors caspase-6, caspase-3 and caspase-7. 
Initiator caspases cleave effector procaspases, activating them~\cite{julien17}. 
Effector caspases are considered the executioners of apoptosis: they cleave structural proteins, leading to a controlled cell demolition.

\begin{figure}[t] \label{fig:network}
\centering
\includegraphics[width = 0.45\textwidth]{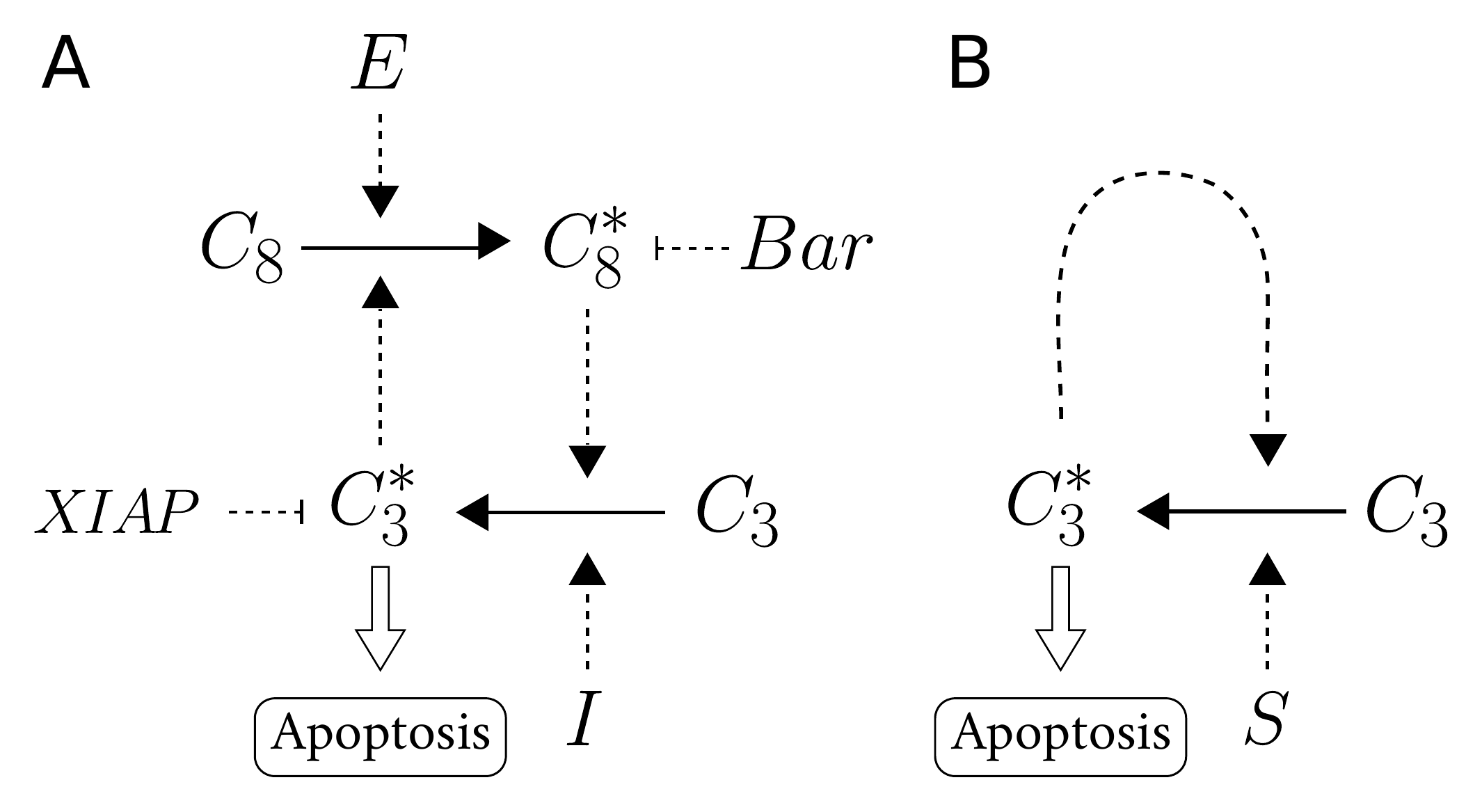}
\caption{
(A) Schematic representation of the core signaling network controlling apoptosis. 
Symbols: $C_{i}$ and $C^{*}_{i}$ stand for the inactive and active forms of caspase $i$, $E$ and $I$ are extrinsic and intrinsic stimuli.
(B) Reduced model centered on caspase-3. The interaction through caspase-8 is represented as an effective positive feedback and $S$ stands for a generic stimulus.
(A, B) Solid arrows indicate caspase activation. Dashed lines indicate indirect activation (arrows) and inhibition (blunt arrows). Hollow arrows indicate apoptotic outcome.
}
\end{figure}

Both extrinsic and intrinsic stimuli can activate these enzymes, eliciting the sequence of events that trigger apoptosis. 
An extrinsic stimulus is the binding of an extracellular ligand, generally from the tumor necrosis factor family, to specific transmembrane death receptors~\cite{kaufmann00}. 
This promotes the assembly of a membrane bound signalling complex that cleaves and activates procaspase-8~\cite{gonzalvez10}. 
On the one hand, caspase-8 directly cleaves and activates effector procaspase-3 and procaspase-7, initiating apoptosis. 
On the other hand, caspase-8 can promote mitochondrial outer membrane permeabilization (MOMP). 
This process results in the release of pro-apoptotic proteins into the cytosol that promotes procaspase-9 activation.
Caspase-9 in turn activates effector procaspase-3/7. 
Thus, both these mechanisms result in the activation of procaspase-3/7 and can be represented as an effective activation of procaspase-3/7 by caspase-8, Fig.~\ref{fig:network}A.
Intrinsic stimuli are of intracellular origin such as cell damage due to $\gamma$ and UV radiation, oxidative stress, oncogene activation, and other events that compromise the cell integrity~\cite{kaufmann00}. 
Intrinsic stimuli promote MOMP, thus activating procaspase-3/7 through caspase-9, and are represented as a direct activation of procaspase-3/7, Fig.~\ref{fig:network}A.
Importantly, effector caspase-3/7 not only cleaves essential cellular substrates but it also activates effector caspase-6 that in turn activates initiator caspase-8, closing a self amplifying positive feedback loop within the enzymatic network~\cite{cowling02}.

To prevent the cell from dying in normal conditions, inhibitory molecules keep the onset of apoptosis in check. 
Proteins inhibiting apoptosis act at several points in the network, binding to active caspases and suppressing their function, Fig.~\ref{fig:network}A. 
The Bifunctional apoptosis regulator (Bar) prevents caspase-8 from activating caspase-3/7, both by sequestering procaspase-8 and by inhibiting MOMP~\cite{zhang00}.
The X-chromosome-linked inhibitor of apoptosis protein (XIAP) binds to caspase-3/7 active site inhibiting its activity and promoting its degradation~\cite{holcik01}.

The complex signaling network controlling apoptosis integrates information from the environment and cell state to take a critical cell fate decision. 
In normal conditions, a certain tolerance to weak stimuli should be possible, to avoid initiating self destruction due to random fluctuations.  
Bistability would allow for this tolerance of the survival state, since only a strong enough stimulus would trigger a switch to the death state.
Additionally, to avoid genomic instability and tumor propagation, commitment to cell death should be irreversible once triggered.
Here we seek an effective description of caspase dynamics that is endowed with these features, bistability and irreversibility.
We focus on caspase-3 as a readout of cell fate since its drastic increase signals the onset of apoptosis.
We integrate intermediate steps of the network into a nonlinear feedback function centered on caspase-3 together with a direct stimulus producing caspase-3 activation, Fig.~\ref{fig:network}B. 
We describe regulatory interactions from Bar and XIAP implicitly through saturating kinetics.
We show that this minimal theoretical description of caspase-3 dynamics harbors a wide bistable region. 
Furthermore, we find that above some critical feedback strength in this region the system can switch irreversibly from survival to death state.
Finally, we use the theory to simulate dynamic stimulation conditions and predict cell fate outcomes in different scenarios. 

\section{Theory}
Caspase-3, an effector caspase at the onset of apoptosis that digests cellular components, is a good indicator of cell state. 
Thus, we focus the theory on this particular enzyme and describe the network interaction as nonlinear feedback and activation functions.
We interpret the steady state of active caspase-3 concentration as the readout of the signaling network: steady high concentrations represent cell death while steady low concentrations represent cell survival.

We propose a two-variable description for the concentrations of the inactive and active caspase-3, $C_3$ and $C_3^*$ respectively,   
\begin{multline} \label{eq:cas3}
\dot{C}_3 = b - d C_3 - k_s S \frac{C_3}{C_3 + K}  -  \psi \frac{{C_3 ^*}^h}{{C_{30} ^*}^h + {C_3 ^*}^h} \frac{C_3}{C_3 + K}        \, ,
\end{multline}
\begin{multline} \label{eq:cas3*}
 \dot{{C}_3^*} =  - d C_3^* + k_s S \frac{C_3}{C_3 + K} +  \psi \frac{{C_3 ^*}^h}{{C_{30}^*}^h + {C_3 ^*}^h} \frac{C_3}{C_3 + K}    \, .
\end{multline}
Inactive caspase-3 is synthesized at a constant basal rate $b$ and degraded at a rate $d$, Eq.~(\ref{eq:cas3}).
Here we assume that active caspase-3 is degraded at the same rate $d$ for simplicity, Eq.~(\ref{eq:cas3*}).
Both extrinsic and intrinsic stimuli promote the activation of different initiator caspases that subsequently leads to effector caspase activation, here represented by caspase-3. 
Thus, in the model we consider a single activation mechanism, represented by a loss term in Eq.~(\ref{eq:cas3}) and a corresponding gain term in Eq.~(\ref{eq:cas3*}).
Caspase-3 activation occurs at a rate $k_s$ modulated by stimulus intensity $S$.
Since conversion of caspase-3 into its active form depends on substrate availability, we assume saturating kinetics in this term.
The functional form of this term is linearly increasing for low caspase concentrations and saturating for large concentrations,  $C_3 \gg K$.
This concentration scale $K$ effectively represents inhibitory mechanisms mediated by proteins such as Bar, Fig.~\ref{fig:network}A.
The last term corresponds to activation of caspase-3 due to the positive feedback from its active form mediated by caspase-6 and caspase-8. 
The feedback strength $\psi$ could be linked to caspase-6 or caspase-8 concentrations, not explicitly included in this description.
This positive feedback comprises several steps and molecular reactions and therefore is here represented by a non-linear function.
The non-linearity takes the form of a sigmoidal Hill function, gradually increasing from zero to one with active caspase-3 concentration. 
The feedback kicks in when active caspase concentration exceeds the threshold $C^*_{30}$, and Hill exponent $h$ defines the steepness of the feedback.
The concentration scale $C^*_{30}$ accounts for regulatory proteins that prevent effector caspase activity, such as XIAP.
A saturating function modulates the feedback term, accounting for substrate depletion as in the stimulus term.

Before considering the steady states of this signaling network, we render the system dimensionless to reduce the number of parameters.
We introduce a concentration scale $C_0$ and a timescale $T$, and define dimensionless variables $x$, $y$ and $t'$:
\begin{equation}
C_3 = C_0 x ,\quad C_3^* = C_0 y \quad \mbox{and} \quad t = T t'  .
\end{equation}
Multiplying Eqs.~(\ref{eq:cas3}) and (\ref{eq:cas3*}) by $T/C_0$ to render them dimensionless, we obtain in terms of dimensionless variables
\begin{multline} %\label{eq:sink_profile}
\frac{\partial x}{\partial t'} = \frac{b T}{C_0} - d T x  - \frac{k_s S T} {C_0} \frac{x}{x + K/C_0} \\ - \frac{\psi T}{C_0}  \frac{y^h}{{C_{30}^*}^h/{C_0}^h + y^h} \frac{x}{x + K/C_0} \, ,
\end{multline}
\begin{multline} %\label{eq:sink_profile}
\frac{\partial y}{\partial t'} =  - d Ty  +  \frac{k_s S T} {C_0} \frac{x}{x + K/C_0} \\ + \frac{\psi T}{C_0}  \frac{y^h}{{C_{30}^*}^h/{C_0}^h + y^h} \frac{x}{x + K/C_0} \, .
\end{multline}
We define dimensionless groups
\begin{equation} \label{eq:dimensionlessgroups}
\beta \equiv \frac{b T}{C_0} , \quad \sigma \equiv  \frac{k_s S T}{C_0},\quad
  \varphi \equiv \frac{\psi T}{C_0}, \quad \kappa \equiv \frac{K}{C_0}, 
\end{equation}
and set the time and concentration scales 
\begin{equation}
T d \equiv 1 \quad \mbox{and} \quad \frac{C_{30}^*}{C_0} \equiv 1  .
\end{equation}
Thus, time is measured in terms of the inverse degradation rate, and concentrations in terms of the feedback activation threshold. 
Renaming the variable $ t' \rightarrow t$ to alleviate notation, we arrive at the dimensionless equations
\begin{equation}
 \dot{x} = \beta - x  - \sigma \frac{x}{x + \kappa} - \varphi \frac{y^h}{1 + y^h} \frac{x}{x + \kappa} \, ,
\label{eq:xdot}
\end{equation}
\begin{equation}
 \dot{y} =  - y  + \sigma \frac{x}{x + \kappa} + \varphi \frac{y^h}{1 + y^h} \frac{x}{x + \kappa} \, ,
\label{eq:ydot}
\end{equation}
where dimensionless parameter 
$\beta$ is the basal synthesis rate, 
$\sigma$ the intensity of the stimulus, 
$\varphi$ the feedback strength, and 
$\kappa$ the relative concentration of inhibitory proteins Bar and XIAP, see Eq.~(\ref{eq:dimensionlessgroups}).

Introducing the total caspase-3 concentration $z = x + y$, from Eqs. (\ref{eq:xdot}) and (\ref{eq:ydot}) we obtain
\begin{equation}
 \dot{z} =  \beta  - z ,
\label{eq:zdot}
\end{equation}
with solution
\begin{equation}
 z(t) = \beta - (\beta - z(0)) e^{-t} .
\label{eq:zsol}
\end{equation}
Thus, the total caspase-3 concentration is decoupled from other variables and has a fixed point $z=\beta$ that is approached exponentially within one unit of dimensionless time.
Here we consider a regime in which the cell is in a stationary state of the network upon receiving different stimuli.
Then, before the stimulus is applied, the total caspase concentration had enough time to settle in this steady state where $z=\beta$ is constant.
In this regime, we can eliminate the inactive caspase-3 concentration described by $x$ in terms of this total caspase-3 concentration,
$x = \beta - y$.
Active caspase-3 dynamics is then described by the single equation
\begin{equation}
 \dot{y} =  - y +  \sigma \frac{\beta -y }{ \kappa + \beta - y} + \varphi \frac{y^h}{1 + y^h} \frac{\beta - y}{ \kappa + \beta - y} .
\label{eq:ydot_1D}
\end{equation}
This dimensionless model contains five parameters: the basal synthesis rate $\beta$ setting the steady state total caspase concentration, the stimulus intensity $\sigma$, the feedback strength $\varphi$, the relative concentration of inhibitory proteins $\kappa$ and 
the exponent of the Hill function $h$ controlling feedback steepness.  

In the following, we analyze whether this model is compatible with properties observed in the biological context such as bistability and irreversibility. 
We construct numerical bifurcation diagrams in Python.
Stable fixed points are obtained as steady states of the dynamics, by numerical integration of Eq.~(\ref{eq:ydot_1D}) with the ODE integrator from the scipy integrate submodule. 
Unstable fixed points are obtained by finding the corresponding root of $\dot y = 0$, using the fsolve function from the scipy optimize submodule~\cite{virtanen20}.

\section{Phase portrait and bifurcation diagram}
\begin{figure}[t!]
\centering
\includegraphics[width = 0.45\textwidth]{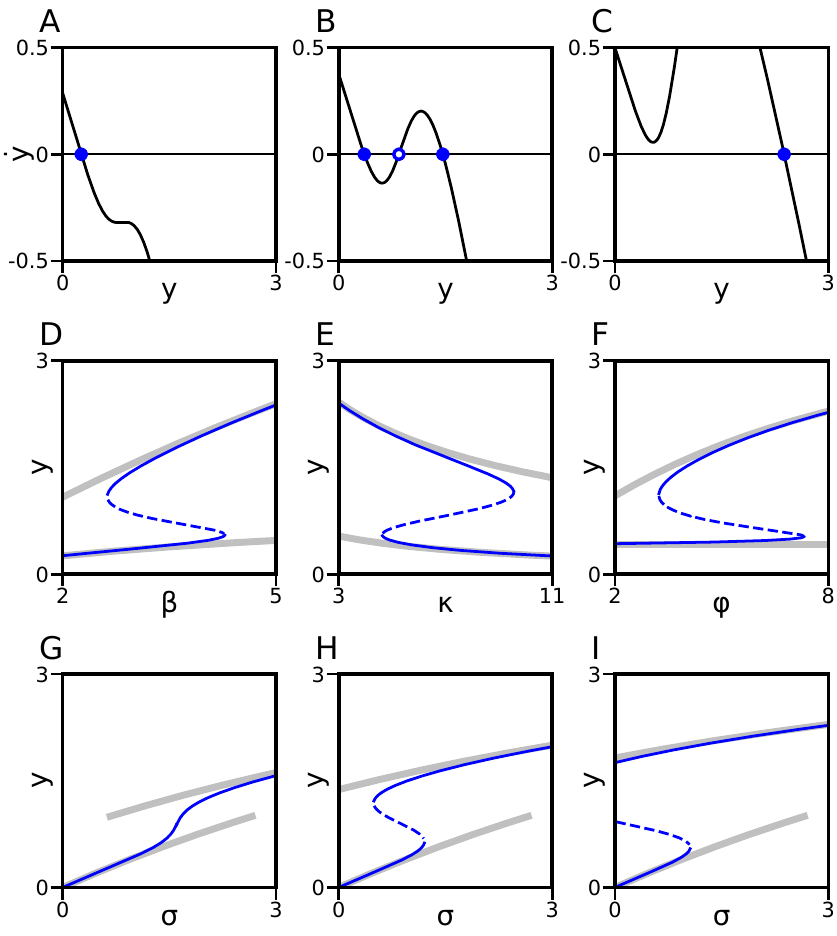}
\caption{Phase portrait and bifurcation diagrams.
(A-C) Phase portrait showing $\dot{y}$ vs. $y$ (solid black line), for three different values of the total caspase concentration $\beta$. 
Stable and unstable fixed points are indicated as full and open blue dots respectively. 
(D-I) Single parameter bifurcation diagrams showing stable (solid blue lines) and unstable (dashed blue lines) fixed points for 
(D) the total caspase concentration $\beta$, 
(E) the relative concentration of inhibitory proteins $\kappa$, 
(F) the feedback strength $\varphi$ and 
(G-I) the stimulus intensity $\sigma$. 
Gray lines are the analytical solutions obtained in the highly nonlinear regime. 
Panels (G-I) correspond to different values of the feedback strength $\varphi$. 
Parameters as in Table~\ref{tab:parametersFig2}.
}
\label{fig:2}
\end{figure}

In this one-dimensional representation, the fixed points of the dynamics are the solutions to $\dot{y}=0$, see Fig.~\ref{fig:2}A-C.
Increasing the parameter $\beta$ we find 
(i) a unique stable fixed point with low active caspase-3 concentration, Fig.~\ref{fig:2}A (ii) two stable fixed points separated by an unstable fixed point, Fig.~\ref{fig:2}B and (iii) one stable fixed point with high active caspase-3 concentration, Fig.~\ref{fig:2}C.
Thus, the model has a built in bistable regime, that is, there are parameter values that allow for two possible steady states, Fig.~\ref{fig:2}B.
We interpret a steady state with low active caspase-3 concentration as a cell survival outcome, and a state with high active caspase-3 concentration as an apoptotic fate. 

In the strongly nonlinear regime $h \gg 1$ it is possible to obtain analytical expressions for fixed points.
In this situation the feedback term $y^h/(1 + y^h)$ approximates a Heaviside step function $H(y-1)$ at $y=1$. 
Although this introduces a discontinuity in Eq.(\ref{eq:ydot_1D}), it also simplifies the solutions to $\dot y = 0$.
For $y>1$ the cell death solution is
\begin{equation}
y_{\rm D} = \frac{1}{2} \left( \beta + \kappa + \sigma + \varphi - \sqrt{\left(\beta + \kappa + \sigma + \varphi\right)^2 - 4\beta(\sigma + \varphi)}\right)
\label{eq:y_d}
\end{equation}
while for $y<1$ the feedback term vanishes and the cell survival solution is
\begin{equation}
y_{\rm S} = \frac{1}{2} \left( \beta + \kappa + \sigma - \sqrt{\left(\beta + \kappa + \sigma \right)^2 - 4\beta \sigma}\right) .
\label{eq:y_s}
\end{equation}
This analysis shows that in the strongly nonlinear regime the system is bistable over wide ranges of parameter values, gray lines in Fig.~\ref{fig:2}D-I. 
Away from bifurcations, the steady states for finite $h$ coincide with the analytical solutions obtained for the strongly nonlinear regime, blue lines in Fig.~\ref{fig:2}D-I.
Near bifurcations the difference with the strongly nonlinear case becomes prominent.
However, the existence of a bistable regime is preserved with moderate nonlinearity.

We are interested in the response of the system to stimuli, here parametrized by the stimulus intensity $\sigma$. 
For low feedback strength $\varphi$, we observe that an increasing stimulus causes a smooth crossover from low to high active caspase concentrations, Fig.~\ref{fig:2}G. 
A larger feedback strength causes a region of bistability in terms of the stimulus intensity $\sigma$, Fig.~\ref{fig:2}H. 
This means that low stimuli will not affect the output of the system significantly, yet increasing the stimulus over some critical threshold can cause an abrupt change of the active caspase concentration and trigger cell death.
Once in this large active caspase concentration state, a moderate reduction in the stimulus does not bring the cell back to the survival state.
However, a strong reduction in the stimulus could still bring this state back for this feedback strength.
For even larger values of the feedback strength the switch becomes irreversible, since even reducing the stimulus back to zero cannot bring back the cell from the death state, Fig.~\ref{fig:2}I. 
Thus, a large feedback strength allows for irreversible apoptotic switch: 
above a critical stimulus intensity the cell goes into the death state without the possibility to return~\cite{tyson03}.
\begin{table}[t]
\centering
\begin{tabular}{|p{1.2cm}|p{1cm}|p{1cm}|p{1cm}|p{1cm}|} 	
 \hline
 \multicolumn{5}{|c|}{Figure \ref{fig:2} parameters} \\
 \hline
 Panel & $\beta$ & $\kappa$ & $\varphi$ & $\sigma$ \\
 \hline
 A  & 2 & 5 & 6 & 1 \\
\hline  
 B  & 3 & 5 & 6 & 1 \\
\hline  
 C  & 5 & 5 & 6 & 1 \\
\hline  
 D  & [2,5] & 5 & 6 & 1 \\
\hline  
 E  & 4 & [3,11] & 6 & 1 \\
\hline   
 F  & 4 & 5 & [2,8] & 1 \\
\hline   
 G  & 4 & 5 & 2 & [0,3] \\
\hline   
 H  & 4 & 5 & 4 & [0,3] \\
\hline   
 I  & 4 & 5 & 6 & [0,3] \\
 \hline
\end{tabular}
\caption{Figure \ref{fig:2} parameters. In all cases $h = 5$.}
\label{tab:parametersFig2}
\end{table}

\section{Fate maps}
\begin{figure}[h]
\centering
\includegraphics[width = 0.45\textwidth]{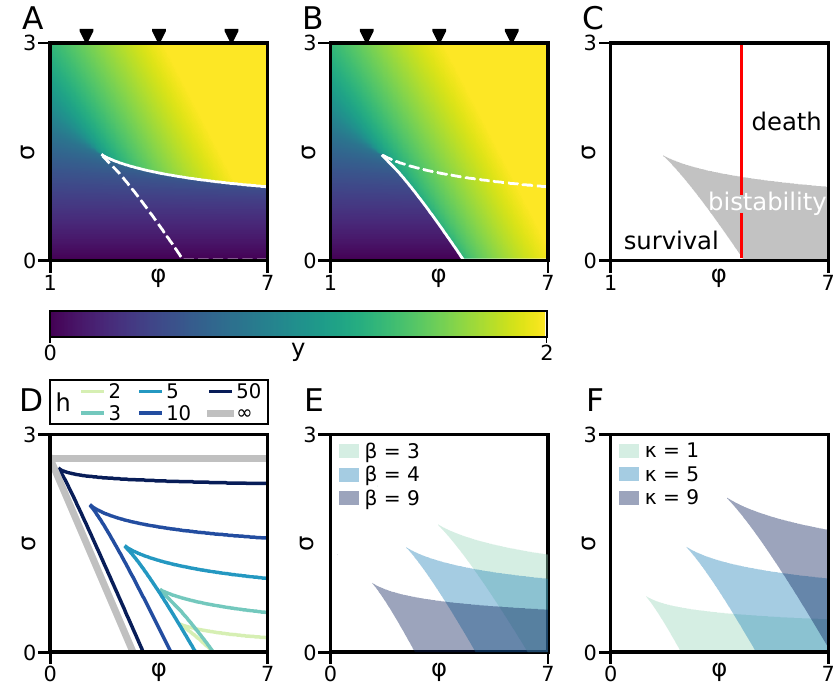}
\caption{Bistability and irreversibility in the two parameter bifurcation diagram for the stimulus intensity $\sigma$ and feedback strength $\varphi$.
(A-B) Stationary value of active caspase concentration $y$ as indicated in the color scale bar. 
White lines delimit the bistable region: solid white lines mark the bifurcations and dashed white lines are shown for reference.
Panels A and B highlight the two possible values of the steady states in the bistable region located in the bottom right corner.
Black triangles mark the feedback strength values from Fig.\ref{fig:2}G-I. 
(C) Cell fate map with bistable region shaded in grey. Vertical red line indicates the irreversibility threshold for feedback strength $\varphi_{\rm i}$.
(D-F) Bistable region for different values of 
(D) the exponent of the feedback Hill function $h$, 
(E) the total caspase concentration $\beta$ and 
(F) the relative concentration of inhibitory proteins $\kappa$.   
(A-F) Unless otherwise specified parameters are: $\beta = 4$, $\kappa = 5$, $h = 5$.}
\label{fig:3}
\end{figure}

We systematically explored the parameters that allow for this kind of behavior constructing two-parameter bifurcation diagrams, Fig.~\ref{fig:3}.
The space defined by the feedback strength and stimulus intensity is divided in three regions: a region where active caspase concentration is low, a region where it is high and a region where both low and high steady state values coexist, Fig.~\ref{fig:3}A,B. 
We can interpret this as a cell fate map with a cell survival region, a cell death region and a region of coexistence where both fates are possible, Fig.~\ref{fig:3}C. 
This bistable region is delimited by two bifurcation curves that converge into a single point forming a cusp.
Beyond a critical feedback strength $\varphi_{\rm i}$ the switch between survival and death becomes irreversible:
after visiting a high caspase concentration state, reducing $\sigma$ from large values cannot cause a transition back from high to low active caspase concentrations.

The bistable region exists even for low nonlinearity and its size increases with nonlinearity $h$, Fig.~\ref{fig:3}D.
In the strongly nonlinear regime $h \gg 1$ we can obtain analytical expressions for the bifurcation curves delimiting the bistable region.
In this regime, fixed points representing death and survival states are given by Eq.~(\ref{eq:y_d}) and Eq.~(\ref{eq:y_s}).
Bifurcation curves mark the boundaries in the fate map where these fixed points cease to exist, at $y=1$.
Thus, setting $y_D = 1$ and $y_S = 1$ we obtain the curves 
\begin{equation}
\sigma = \frac{\kappa}{\beta -1 } + 1 - \varphi 
\quad \mbox{and} \quad
\sigma = \frac{\kappa}{\beta -1 } + 1 \, ,
\end{equation}
thick grey lines in Fig.~\ref{fig:3}D.
Increasing the dimensionless synthesis rate $\beta$ shifts the bistable region towards smaller values of $\sigma$ and $\varphi$, Fig.~\ref{fig:3}E.
The total amount of caspase-3 in steady state is $z=\beta$.
In the absence of stimuli all caspase-3 is in its inactive form, $x=\beta$ and $y=0$. 
Thus, in the absence of stimuli inactive caspase levels are larger for larger $\beta$, constituting a large pool of enzyme waiting to be converted into the active form. 
Increasing the relative inhibitor concentration $\kappa$ shifts the bistable region towards larger values of $\sigma$ and $\varphi$, Fig.~\ref{fig:3}F.
The factor $x/(\kappa + x)$ modulates the stimulus of intensity $\sigma$ for converting caspase into the active form, Eq.~(\ref{eq:xdot}).
For $x < \kappa$ the conversion rate factor is smaller than $1/2$, and a larger stimulus intensity $\sigma$ is required to trigger cell death. 
For $\kappa < x$ the conversion rate factor approaches $1$ and a smaller stimulus intensity $\sigma$ is enough to trigger cell death. %, Fig.~\ref{fig:3}F.
Thus, the relative values of $\beta$ and $\kappa$ determine how fast a stimulus will grow.

We have shown that the system displays a robust bistable regime over a wide range of parameter values.
Such bistability means that small stimuli will not significantly change the concentration of active caspase, ensuring the robustness of the survival state.
A large stimulus will push the system over the threshold and trigger apoptosis through the conversion to active caspase that elicits the cell death response.
There is also a wide region in parameter space where the system is irreversible and cannot switch back from high to low active caspase concentrations when the stimulus is reduced.
We argue that irreversibility is a natural requirement from the biological perspective, since apoptosis involves processes that damage cell components permanently. Thus, once apoptosis is initiated there should not be a way back since its arrest would leave the cell nonfunctional.
%
%
%
\begin{comment}
\begin{table}[b]
\centering
\begin{tabular}{|p{1.2cm}|p{1.2cm}|p{1.2cm}|p{1.2cm}|p{1.2cm}|p{1.2cm}|} 	
 \hline
 \multicolumn{6}{|c|}{Figure \ref{fig:3} parameters} \\
 \hline
 Panel & $\beta$ & $\sigma$ & $\kappa$ & $\varphi$ & $h$ \\
 \hline
 A  & 4 & - & 5 & - & 5\\
\hline  
 B  & 4 & - & 5 & - & 5\\
\hline  
 C  & 4 & - & 5 & - & 5\\
\hline  
 D  & 4 & - & 5 & - & -\\
\hline  
E  & [3,5,9] & - & 5 & - & 5\\
\hline  
 F  & 4 & - & [1,5,9] & - & 5\\
\hline
\end{tabular}
\caption{Figure \ref{fig:3} parameters.}
\label{tab:parametersFig3}
\end{table}
\end{comment}

\section{Time-dependent stimuli}
We have considered the role of the feedback in generating a bistable regime and irreversibility under constant stimulation conditions.
As discussed in the introduction, apoptosis can be triggered both by external stimuli such as the binding of ligands to death receptors, or by internal stimuli produced by cell states that result from damage to cell components.
In both cases, the stimulus can be time dependent rather than permanent.
Motivated by these physiologically relevant situations, we are interested in analyzing the response of the system upon transient stimulation regimes. 
To incorporate such transient stimulation in the theory we introduce a time dependence in the parameter controlling stimuli intensity.

Cells may receive transient strong signals that are meant to induce apoptosis, for example delivered from the surrounding tissue during development.
However, weaker confounding signals may also reach the cell transiently, and in these cases the cell should avoid entering apoptosis. 
Given the drastic outcome of apoptosis, the apoptotic network should distinguish these situations.

\begin{figure}[h]
\centering
\includegraphics[width = 0.45\textwidth]{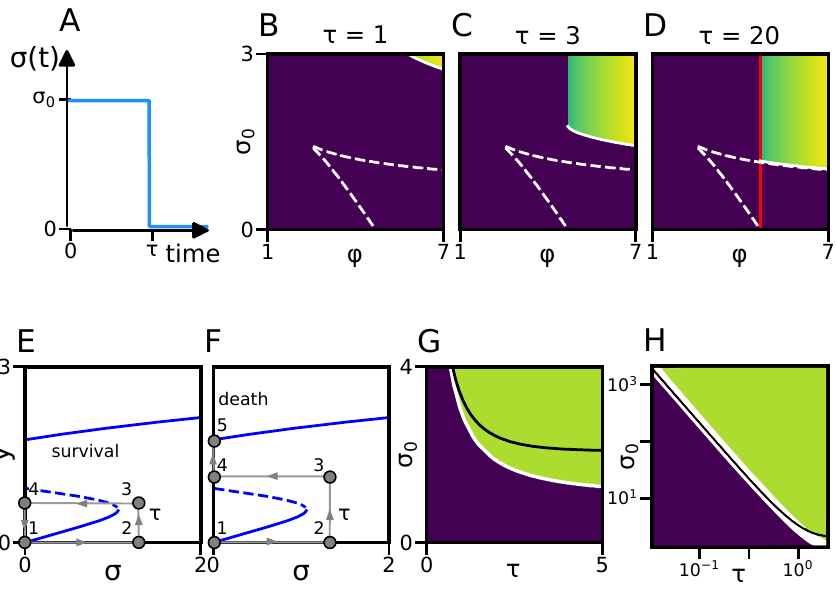}
\caption{Transient stimulation. 
(A) Time dependent stimulus of intensity $\sigma_0$ and duration $\tau$.
(B-D) Cell fate maps for the stimulus intensity $\sigma_0$ and feedback strength $\varphi$, with increasing stimulus duration $\tau$ as indicated. 
Solid white line marks the critical lower boundary of the cell death region, $\sigma_c(\varphi)$.
Dashed white line indicates the bistable region for reference. 
Red vertical line in (D) indicates the feedback threshold $\varphi_{\rm i}$ for irreversibility.
(E, F) Schematic trajectories in the steady state bifurcation diagram for stimulus intensity $\sigma$, during transient perturbations that finish in (E) cell survival and (F) cell death.
In both cases, the system is initially in the survival steady state $y=0$ (1), and a stimulus of intensity $\sigma_0$ is applied (2). While the stimulus lasts, active caspase concentration $y$ increases (3). The stimulus is removed after the time $\tau$ (4). If the value of $y$ has exceeded the value of the unstable fixed point for $\sigma = 0$ (see dashed blue line) $y$ keeps increasing up to the apoptotic steady state value, (5) in panel (F). If not, it falls back to the survival steady state $y=0$ in panel (E).
(G, H) Cell fate map for the stimulus intensity $\sigma_0$ and duration $\tau$, for a feedback strength $\varphi = 6$, in (G) linear and (H) log-log scales.
White line indicates the critical curve that separates the survival state (violet) from the death state (green). 
Black line indicates the critical curve obtained analytically for short stimuli, Eq.~(\ref{eq:solshorttimes}). 
(B-D,G,H) Color indicates the stationary value of active caspase concentration $y$ with scale as in Fig.~\ref{fig:3}. 
Parameters values are $\beta = 4$, $\kappa = 5$, and $h = 5$.
}
\label{fig:4}
\end{figure}

Thus, we first examine the conditions for a finite stimulus to be lethal for the cell.
We consider a situation where a cell is in the survival steady state under no stimulation and receives a transient stimulus. 
For simplicity, here we consider a square-pulse stimulus, that is a transient stimulus $\sigma (t)$ of finite duration $\tau$ and intensity $\sigma_0$, Fig.~\ref{fig:4}A.
We evaluate the outcome of such a finite stimulus for different feedback strengths and stimulus intensities, and for varying stimulus duration.
We find that there is a threshold in the feedback strength to enter the apoptotic fate region, Fig.~\ref{fig:4}B-D. 
This threshold value coincides with the minimum feedback strength $\varphi_{\rm i}$ that renders the system irreversible. 
Because the stimulus is eventually withdrawn ($\sigma = 0$ for $t > \tau$), active caspase concentration will only remain in the apoptotic state in the cases where it is impossible to return to the cell survival state, Fig.~\ref{fig:2}I. 
For feedback strength $\varphi$ larger than this critical value, apoptosis still depends on the intensity and duration of the stimulus.
If the intensity of the stimulus falls below the upper boundary of the cusp, active caspase concentration never overcomes the value of the lower stable fixed point and returns to $y=0$ when the stimulus is removed.
Above the cusp, there is a critical line $\sigma_c (\varphi)$ setting a minimum stimulus intensity to enter apoptosis, depending on the feedback strength.
This critical line depends on stimulus duration $\tau$, Fig.~\ref{fig:4}B-D.
This is because to enter apoptosis, active caspase concentration is required to surpass the value $y^*$ of the unstable fixed point for $\sigma = 0$, such that when the stimulus is removed the system is on the death state attraction basin, Fig.~\ref{fig:4}E,F. 
Thus, there is a critical curve that separates the survival state from the death state in the fate map for stimulus intensity and duration, Fig.~\ref{fig:4}G, H.
We can approximate the dynamics for small $y$ to obtain a curve $\sigma(\tau)$ for short stimuli.
In the absence of stimuli $\sigma = 0$, the steady state is $x = \beta$ with $y = 0$ and the feedback term is off in Eq.~(\ref{eq:ydot_1D}).
When a stimulus of intensity $\sigma$ is applied active caspase dynamics for short times can be approximated by
\begin{equation}
\dot{y} \approx  -y + \sigma \frac{\beta }{\beta +\kappa} \, .
\label{eq:shortime}
\end{equation}
We can readily integrate Eq.~(\ref{eq:shortime}) and impose the condition $y(\tau)=y^*$ to obtain an approximate critical curve
\begin{equation}
\label{eq:solshorttimes}
\sigma(\tau) =  y^* \frac{\beta +\kappa}{\beta}\frac{1}{1 -e^{-\tau}} \, ,
\end{equation}
Fig.~\ref{fig:4}H.
This analytical insight indicates that there is no critical duration, since for a very short stimulus there is always a value of stimulus intensity that triggers apoptosis.

\begin{figure}[t]
\centering
\includegraphics[width = 0.45\textwidth]{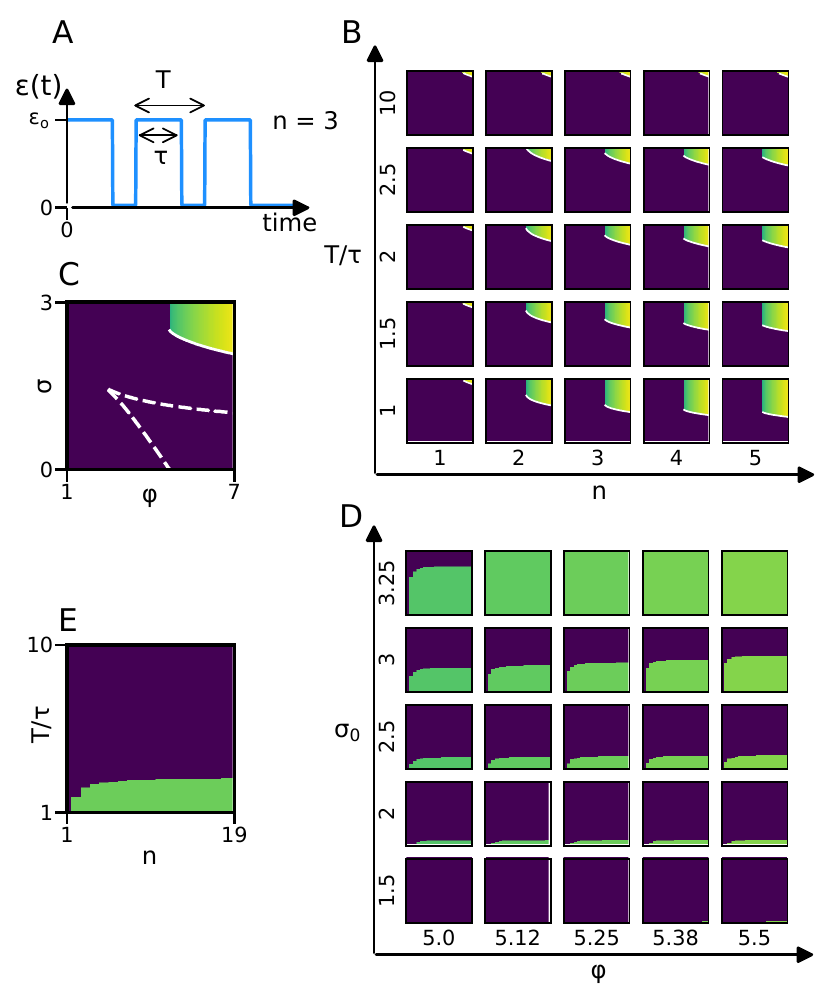}
\caption{Pulsed stimulation.
(A) Time dependent stimulus is a train of $n$ pulses of intensity $\sigma_0$ and duration $\tau$, separated a time $T$. 
(B) Cell fate maps for the stimulus intensity $\sigma_0$ and feedback strength $\varphi$ as in Fig.~\ref{fig:4}(B-D), for different number of pulses $n$ and $T/\tau$ values. 
(C) Enlarged is the cell fate map for a stimulus of $n = 3$ pulses with $T/\tau = 2$. 
(D) Cell fate maps for $T/\tau$ and the number of pulses $n$ for different values of stimulus intensity $\sigma_0$ and feedback strength $\varphi$.
(E) Enlarged is the cell fate map for a stimulus intensity of $\sigma_0 = 2.5$ and a feedback strength $\varphi = 5.25$. 
%
% In panels (B, D, E) the separation time is scaled by the pulse duration $\tau$.
%
In all cases $\tau = 1$, $\beta = 4$, $\kappa = 5$, $h = 5$. 
Color scale as in Fig.~\ref{fig:3}.
}
\label{fig:5}
\end{figure}

Beyond single pulses, cells may receive signals with more complex temporal patterns.
Here we consider pulse trains, which provide a systematic way to probe the response to frequency and duration, and could be realized experimentally.
We introduce a time-dependent stimulus with a number $n$ of pulses of intensity $\sigma_0$ and duration $\tau$, that are separated by a time interval $T$, Fig.~\ref{fig:5}A.
We explore the conditions on the number of pulses $n$ and pulse separation $T$ to trigger apoptosis. 
We first map cell fate in terms of feedback strength and stimulus intensity, for pulse trains with different pulse separation and number of pulses, Fig.~\ref{fig:5}B, C.
As we observed for a single pulse, there is a minimum feedback strength required to enter apoptosis, coinciding with the minimum feedback strength needed for the system to be irreversible. 
When the separation interval between pulses equals the pulse duration, increasing the number of pulses is equivalent to a single pulse of increasing duration, lower row in Fig.~\ref{fig:5}B.
For a fixed number of pulses, increasing the pulse separation results in an increasing minimum stimulus intensity for entering apoptosis, columns in Fig.~\ref{fig:5}B.
This is because in between pulses the active caspase concentration drops transiently before the next pulse arrives.
For a fixed pulse separation, increasing the number of pulses in the train reduces the minimum stimulus intensity required to enter apoptosis, rows in Fig.~\ref{fig:5}B.
As pulse separation keeps increasing, the pulses become effectively isolated and the fate maps are identical to that of a single pulse, top row in Fig.~\ref{fig:5}B.
The changes observed in the minimum stimulus intensity required for apoptosis prompted us to map cell fate in terms of the pulse separation and their number for constant feedback and stimulus intensity, Fig.~\ref{fig:5}D, E.
In these maps we observe that for pulse separation larger than a critical value the cell survival fate prevails independently of the number of stimuli pulses, Fig.~\ref{fig:5}E. 
Both increasing feedback strength and stimulus intensity results in an increase of this critical time separation, Fig.~\ref{fig:5}D.
Taken together, these fate maps reflect the activation dynamics during a pulse train: if the time interval between pulses is larger than the time it takes to return to the survival state, there is no cumulative effect due to consecutive pulses and the cell may survive. 
Activation dynamics depends on stimuli intensity and feedback strength and for that reason so does the critical time interval, Fig.~\ref{fig:5}E. 

\section{Discussion}
In this work, we present a low dimensional description of the apoptotic switch focused on caspase-3 dynamics.
This description integrates intermediate events in the network into a nonlinear positive feedback function acting on effector caspase-3, a readout of cell fate.
We integrate stimuli that trigger apoptosis and inhibitory elements into a saturating activation function. 
We find that the theory has a robust bistable regime where both survival and death are possible cell fates. 
In this regime, the system tolerates caspase-3 concentration perturbations, and has a sharp transition to cell death when a threshold is exceeded.  
For a range of parameter values, this switch to the death fate is irreversible.
This may describe commitment to cell death once a critical level of cell damage has been traversed. 
We use this theory to explore the response to time-dependent stimuli.
For a single pulse stimulus we find a critical feedback strength that allows for such transient stimuli to permanently trigger cell death.
While very short stimuli can trigger apoptosis if their intensity is large enough, there is a critical stimulus intensity required to trigger apoptosis, even for long stimuli duration.
Finally, we consider stimulation with pulse trains and find that multiple consecutive pulses can trigger apoptosis if the separation between pulses is short enough.

Previous theoretical descriptions have focused on different aspects of the signaling network controlling apoptosis~\cite{eissing04, bagci06, legewie06, chen07, cui08, zhang09}.
The effective mutual activation between caspase-3 and -8 was explicitly described together with inhibitors Bar and XIAP~\cite{eissing04}.
This work suggested that control at the level of initiator caspases was key to generate a bistable regime compatible with physiological parameter values.
Active caspase-3 is tagged by XIAP for ubiquitination, adding an additional degradation mechanism that might result in an effective degradation rate larger than that of inactive caspase-3~\cite{holcik01}. 
While this differential degradation has been contemplated in some works~\cite{eissing04}, here we consider the same degradation rate for active and inactive caspase for simplicity. 
In our theory, such differential degradation would couple total caspase concentration to the active caspase and therefore change with perturbations from stimuli.
It will be interesting to study how this affects steady states and cell responses to stimuli.

Bistability has been also addressed in the context of the mitochondrial sector of the network~\cite{bagci06,legewie06,chen07,cui08}.
Mitochondrial outer membrane permeabilization (MOMP) is tightly regulated by the balance of pro- and anti-apoptotic proteins in the Bcl-2 family~\cite{penablanco17, singh19}. 
Downstream of MOMP, caspase-9 molecules assemble into the apoptosome, where they self activate~\cite{srinivasula98}. 
One study proposed that cooperative apoptosome assembly may be a robust mechanism for inducing bistability~\cite{bagci06}.
Another study focused on mutual activation between caspase-9 and -3, and their inhibition by XIAP, which establishes an implicit positive feedback that promotes bistability together with irreversibility~\cite{legewie06}.
Further works proposed that bistability and irreversibility may arise upstream from mitochondria, as a consequence of interactions among proteins that regulate MOMP~\cite{chen07, cui08, sun10}.
In subsequent work, these different aspects of the complex molecular network controlling apoptosis were organized into three modules: initiator, amplifier and executioner~\cite{zhang09}.
The initiator included interactions among the Bcl-2 family proteins, the amplifier described MOMP and the release of pro-apoptotic components, and the executioner module covered apoptosome formation, and activation of caspase-9 and -3. 
This work suggested that positive feedback in the initiator module may cause bistability –and a sharp activation, while positive feedback in the executioner module may be responsible for irreversible activation.
In our effective description, we consider a single feedback loop between caspase-8 and caspase-3, possibly mediated by caspase-6.
Feedback is responsible for bistability, and further increasing feedback strength may also cause irreversibility.
Alternatively, irreversibility may be induced both by increasing the total caspase concentration here parametrized by $\beta$, or by decreasing the relative concentration of inhibitory proteins $\kappa = K/C^*_{30}$, such as Bar and XIAP.

In recent years, experimental methods have allowed to study apoptosis at the single cell level, imaging reporters for caspase-3 and caspase-8 substrates, and for MOMP~\cite{albeck08a}.
Individual dynamics revealed a snap-action switch like behavior as well as a variable delay in activation time after stimuli~\cite{albeck08a, albeck08b}. 
The approach was used to study variability in both the timing of extrinsic apoptosis and the outcome of cell fate, in a population of genetically identical cells~\cite{spencer09}.
This variability was described in the model by random sampling the initial conditions from experimentally determined probability distributions~\cite{gaudet12}.
While random initial conditions may account for extrinsic noise, the effects of intrinsic noise have also been considered in stochastic simulations~\cite{ooi13}.
Here, we have considered a deterministic description of caspase-3 dynamics. 
Future work could address the effects of both extrinsic fluctuations –by introducing distributions for parameter values, and intrinsic fluctuations –as an additional dynamic noise term.

Apoptosis is an attractive model system to study fate decisions, with a neatly defined outcome and a solid background understanding of its regulatory network~\cite{taylor08, bedoui20}.
However, some key questions remain open~\cite{green19}, and it has been suggested that the field is poised to integrate theory and experiment to bring insight into the mechanisms that regulate and trigger apoptosis~\cite{tyson07}.
The effective description we propose here has the advantage to provide analytic insights into these mechanisms, and makes some distinct predictions.
One of the outstanding questions in the field is how commitment to cell death is regulated and to what extent a cell may refrain from death once engaged in apoptotic events~\cite{green19}.
Here we showed that in the framework of the theory there is a critical feedback strength necessary for a cell to irreversibly adopt a death fate. 
This critical feedback strength may be regulated both by the relative concentration of inhibitory proteins and by total caspase-3 concentration.
In the theory, feedback strength is an effective quantity resulting from intermediate network components, such as caspase-6.
Thus, interfering with caspase-6 function could be one way to bring the cell below the critical feedback strength required for irreversibility.
This could be tested using transient stimulation and observing the nature of cell response.
For transient stimuli, we showed that there is a critical stimulus intensity to trigger apoptosis.
The model predicts the shape of the critical curve between the duration and intensity of transient stimuli that trigger apoptosis.
For pulsed stimulation, the model predicts a characteristic relation between the number of pulses and the time interval between them to trigger apoptosis. 
These predictions may be tested imaging single cells during apoptosis~\cite{albeck08a, spencer09, corbat18}.
We expect that our model may be a useful tool to interrogate the dynamics of apoptosis and its underpinnings in constant and variable experimental conditions.

\vspace{0.5cm}

\section*{Acknowledgements}
We thank F. Fabris, D. M. Arribas and M. Wappner for providing valuable feedback on the manuscript. This research was sponsored by ANPCyT PICT 2013 1301 to HEG and LGM, PICT 2017 3753 to LGM, and FOCEM-Mercosur (COF 03/11) to IBioBA.

\end{document}